\documentclass[12pt,preprint]{aastex}

\newcommand{\RSUN}{R$_\sun$}

\slugcomment{Accepted by ApJ 2011 April 28}

\shorttitle{}
\shortauthors{}

\begin{document}

%% LaTeX will automatically break titles if they run longer than
%% one line. However, you may use \\ to force a line break if you desire.

\title{Active Longitudes Revealed by Large-scale  and Long-lived Coronal Streamers}

\author{Jing Li}
\email{jli@igpp.ucla.edu}
\affil{Institute for Geophysics and Planetary Physics, University of California at Los Angeles, \\
Los Angeles, CA  90095-1567}

%% Notice that each of these authors has alternate affiliations, which
%% are identified by the \altaffilmark after each name.  Specify alternate
%% affiliation information with \altaffiltext, with one command per each
%% affiliation.
%% Mark off your abstract in the ``abstract'' environment. In the manuscript
%% style, abstract will output a Received/Accepted line after the
%% title and affiliation information. No date will appear since the author
%% does not have this information. The dates will be filled in by the
%% editorial office after submission.

\begin{abstract}
We use time-series ultraviolet full sun images to construct limb-synoptic maps of the Sun.  On these maps, large-scale, long-lived coronal streamers appear as repetitive sinusoid-like arcs projected over the polar regions.  They are caused by high altitude plasma produced from sunspot-rich regions at latitudes generally far from the poles.  The non-uniform longitudinal distribution of these reveals four longitudinal zones at the surface of the sun from which sunspots erupt preferentially over the 5-year observing interval (2006 January to 2011 April).  Spots in these zones (or ``clusters'') have individual lifetimes short compared to the lifetimes of the coronal features which they sustain, and erupt at different times. The four sunspot clusters contain $> 75$\% of all numbered sunspots in this period. They occupy two distinct longitudinal zones separated by $\sim180^\circ$ and each spanning $\sim100^\circ$ in longitude. The rotation rates of the spot clusters are  $\sim$5\% faster than the rates at both the surface and the bottom of the convection zone. While no convincing theoretical framework exists to interpret the sunspot clusters in the longitude-time space, their persistent and nonuniform distribution  indicates long-lived, azimuthal structures beneath the surface, and are compatible with the existence of previously-reported active longitudes on the sun. 
%The results imply that the global solar magnetic field is non-axisymmetric, the torsional oscillation governs the sunspot zones.
\end{abstract}

% The different journals have different requirements for keywords.  The
% keywords.apj file, found on aas.org in the pubs/aastex-misc directory, 
% contains a list of keywords used with the ApJ and Letters.  These are 
% usually assigned by the editor, but authors may include them in their 
% manuscripts if they wish. 


\section{Introduction}
While it may appear locally chaotic, the Sun's activity is well known to be globally ordered and shows a number of distinct patterns.  Summarized by \citet{1961ApJ...133..572B}, observations of sunspots reveal Hale's law of hemispheric polarity \citep{1919ApJ....49..153H}, Joy's law of polarity pair orientation, and Sp\"{o}rer's law of zone as well as Maunder's sunspot butterfly diagram through the 11-year solar activity cycle, and 22-year magnetic cycle. These well established observations converge with recent discoveries on surface flows, notably the torsional oscillation and the meridional flow \citep{1980ApJ...239L..33H,1984SoPh...92....1D,1996ApJ...460.1027H,2002ApJ...577L..53W,2002Sci...296..101V,2010Sci...327.1350H}. These observations laid foundations for modeling solar global magnetic fields from the Babcock-Leighton's schematic dynamo model \citep{1961ApJ...133..572B,1969ApJ...156....1L} to recent more advanced models \citep{1991ApJ...383..431W,1999ApJ...518..508D,2010LRSP....7....3C}.  An ultimate goal is to understand the rise and fall of the global magnetic field and so to predict solar cycles.

On other stars, the existence of long-lived active longitudes (regions prone to produce an excess number of spots) is well established \citep{1998A&A...336L..25B}.   Active longitudes have been reported on the Sun for many decades, but their existence is not without controversy and they do not find an explanation in standard models of the solar dynamo. The tendency of new sunspot groups to emerge in the neighborhood of previously existing sunspots for several consecutive solar rotations was observed, especially during the solar activity minimum \citep{1994ApJ...422..883G,1998A&A...332..353G,2003ApJ...586..579V}. So-called ``complexes of activity'' on the sun were reported from a statistical study of the sunspot distribution over 128 years and from synoptic maps of photospheric magnetic fields \citep{bogart1982,1983ApJ...265.1056G,2003A&A...405.1121B,2009Ge&Ae..49..866O}. Solar flares, which occur mostly in association with active regions, are found to concentrate on ``active zones"  similar to the behavior of sunspots  \citep{1987ApJ...314..795B}. The analyses of high-flare-activity sunspot regions led to the existence of two ``giant longitude zones'' with one of them flare-richer than the other \citep{1989SoPh..123...69S,1993SoPh..144..399R}. Analysis of GOES X-ray solar flare data lead to a similar conclusion, with two active longitudinal zones separated by $180^\circ$ (\citet{2011JASTP..73..258Z}).  ``Sunspot nests'' or ``sunspot clusters'' were also found by a {\it single-linkage clustering technique} used to trace the photospheric sunspots in longitude, latitude and time \citep{1990SoPh..129..221B}.
%The phenomenon is important because 
%By definition, the phenomenon is also called the ``sunspot nest'' or ``sunspot cluster'' when sunspots emerge in a confined longitudinal range, but at different time.  This study is carried by a {\it single-linkage clustering technique} to trace the photospheric sunspots in longitude, latitude and time \citep{1990SoPh..129..221B}. Despite repeated reports of the active longitudes, the concept is not fully established in part because of the solar differential rotation. The complicated patterns of the solar rotation result in the difficulty of establishing the phenomenon for long period of time.  Most techniques require more accurate solar rotation rate measurements \citep{1969SoPh....7..210V}.   **JING THIS PART SEEMED TO HANG AND TO BE UNIMPORTANT SO I COMMENTED IT OUT FOR NOW.  WHAT IS THE POINT OF THIS LAST PART?**.

In this work, we use long-lived coronal streamers linked to underlying active regions as a proxy by which to study the longitudinal distribution of sunspot groups on the Sun. Benefiting from a large amount of uninterrupted coronal emission data on the entire sun from space, we demonstrate that the large-scale, long-lived coronal streamers directly reflect ``sunspot active longitudes'' in the photosphere. 
%These coronal streamers appear as repetitive, sinusoidal features crossing dark coronal holes on so-called {\it coronal limb synoptic maps} (LSM) (e.g. Fig. \ref{lsm}) 
%JING - FIGURE 3 CANNOT BE THE FIRST FIGURE.  THEY HAVE TO BE CALLED IN ORDER.
%which are constructed from time-series coronal images. 
In an earlier study, close inspection and a schematic model showed that these streamers are connected with more equatorial active regions \citep{2000ApJ...539L..67L}. During the period April 1996 to May 1997 ($\sim 1$ year), three distinct coronal streamers with life spans as long as 10 solar rotations were found to be sustained by three distinct regions from where sunspots emerged.  The sunspot groups erupt from a localized region of the sun, but at different times \citep{2002ApJ...565.1289L}. %Nevertheless, the existence of active longitude was not drawn from that study because of the short time coverage.
%During the last solar minimum (1996-1997), large-scale, long-lived coronal streamers were clearly linked with sunspot active regions in the photosphere \citep{2002ApJ...565.1289L}. 
% Sunspots in these clusters erupt in the same localized region of the sun, but not at the same time.  
The current study has more than 5 year (January 2006-April 2011) time coverage (compared with 1 year in the previous study) and during this period, the solar activity passed from the declining phase to the minimum and to  the ascending phase. 
%We found that large, long-lived persistent features in the solar corona are connected with the photospheric active regions. In this context, the knowledge of accurate solar differential rotation is not crucial for our technique because  
The new observations point to the little-studied, non-axisymmetric nature of the global solar magnetic field \citep{1971A&A....13..203S,1993A&A...272..621D,2005ApJ...635L.193D,2006A&A...445..703B}. 

%The cycles of activity are also reflected in the solar corona. This is clearly shown by full coronal limb-synoptic maps (LSM) constructed from images of the coronal emission lines \citep{2000ApJ...539L..67L}. Such maps provide a convenient way to visualize solar imaging data over long timescales. They show sunspot regions as bright patches moving from high latitudes to equatorial belt as the cycle progresses from the start to the end. They show that coronal holes are clearly visible during the solar minimum, but  are covered by the high latitude streamers during the solar activity maximum. They show long-lived coronal streamers, associated with active regions, in the form of sinusoidal patterns over the polar holes \citep{2002mwoc.conf..333L}. The strength of these patterns are correlated with the activity cycle. 

%In this work, we examine the distribution of active regions over a 5 year period using images of EUV emission from confined plasma in coronal streamers as a marker.   The observational data consist of EUV images taken by SOHO/EIT \citep{1995SoPh..162..291D}  and STEREO A/SECCHI/EUVI \citep{2008SSRv..136...67H}.  T
\section{Data}
Two sets of data are employed in this study:  full sun coronal EUV images and photospheric sunspot records.

The full sun images are obtained by EUV imagers from space, which record the FeXII 195 \AA~line, for which the effective plasma temperature is $\sim$1.5 MK. As a result, coronal emissions are more prominent in the active regions than in coronal holes and the quiet Sun, where the temperature is normally below 1.5 MK \citep{2007ApJ...659.1673A}. For the period between 2006 January 1 and 2006 December 3, we use images taken by the Extreme ultraviolet Imaging Telescope (EIT) \citep{1995SoPh..162..291D} onboard SoHO \citep{1995SoPh..162....1D}. For the period after 2006 December 4, we use images taken by  the Extreme UltraViolet Imager (EUVI) \citep{2008SSRv..136...67H}, a part of the Sun Earth Connection Coronal and Heliospheric Investigation (SECCHI) package on board STEREO \citep{2008SSRv..136....5K}. 

The STEREO spacecraft are separated from the Earth by a large and time-dependent difference in ecliptic longitude.  To compare sunspot records obtained from the Earth with the EUVI/STEREO images, this longitude difference  has to be taken into account.  For this, we use the World Coordinate System recorded in the STEREO FITS headers \citep{2006A&A...449..791T, 2010SoPh..261..215T}. The spacecraft heliographic longitudes are subtracted from the sunspot disk longitudes as if the sunspots are viewed from STEREO A. For example, the sunspot group NOAA 11045 was reported at W17N23 on 2010 February 9 as viewed from the Earth, but its disk location was E47N23 after correction to the location of STEREO A.  The sunspot disk locations are corrected in this way when the coronal images taken with STEREO A and B are used, but no correction was made when SoHO/EIT data are used, since SoHO's longitude remains close to that of the Earth. 

The photospheric sunspot daily records known as the {\it Solar Region Summary} (SRS) are compiled by the {\it Space Weather Prediction Center} (SWPC, http://www.swpc.noaa.gov/) from approximately a dozen observatories reporting observations in real time. In the SRS records, each sunspot group is assigned a number, known as the NOAA number, during its disk passage. Also included are observing dates, disk locations, Carrington longitudes and latitudes of the sunspots, the longitudinal extents of groups in heliographic degrees, the total corrected sunspot group area in millionths  of the solar hemisphere (equal to 3$\times$10$^6$ km$^2$), the total number of sunspots visible in the group, and a magnetic classification of the group. 

For convenience, the time is measured as Day of Year (DOY), where we define DOY=1 as  2006 January 1.

\section{Large-scale, Long-lived Coronal Streamers}
Our findings are based on the study of the large-scale, and long-lived coronal streamers in the EUV coronal emission data. These streamers are found to be largely associated with more equatorial active regions, but are often seen at high latitudes due to projection effect \citep{2000ApJ...539L..67L}.

\subsection{Large-scale Coronal Streamers}
On individual images, large coronal streamers extending from active regions are best observed at soft X-ray wavelengths (c.f.~Figure 2 of \citet{2002ApJ...565.1289L} or the movie at URL http://www2.ess.ucla.edu/~jingli/movie\_sxt/AlMg\_19960727-19960921.html). This is because these streamers have temperatures $\sim 2.0$ MK (e.g.~\cite{1998ApJ...506..431L}), which are well-matched to the effective temperature range sampled by the imagers on Yohkoh/SXT \citep{1991SoPh..136...37T}.  
%In comparison, the coronal images in EUV wavelengths have less contrasts between hot active regions, their associated streamers and the cold quiet sun and the coronal holes than those in X-ray. %JING _ SOMETHING IS WRONG WITH THE PREVIOUS SENTENCE 
 %Four image panels show, respectively, active regions on the east limb, panel (a); on the disk, panel (b); on the west limb, panel (c); and at the back of the sun, panel (d). 

\begin{figure}[t]
%\epsscale{1.2}
\begin{center}
%\plotone{lca2006339-480noaa.pdf}
\includegraphics[width=1.0\textwidth]{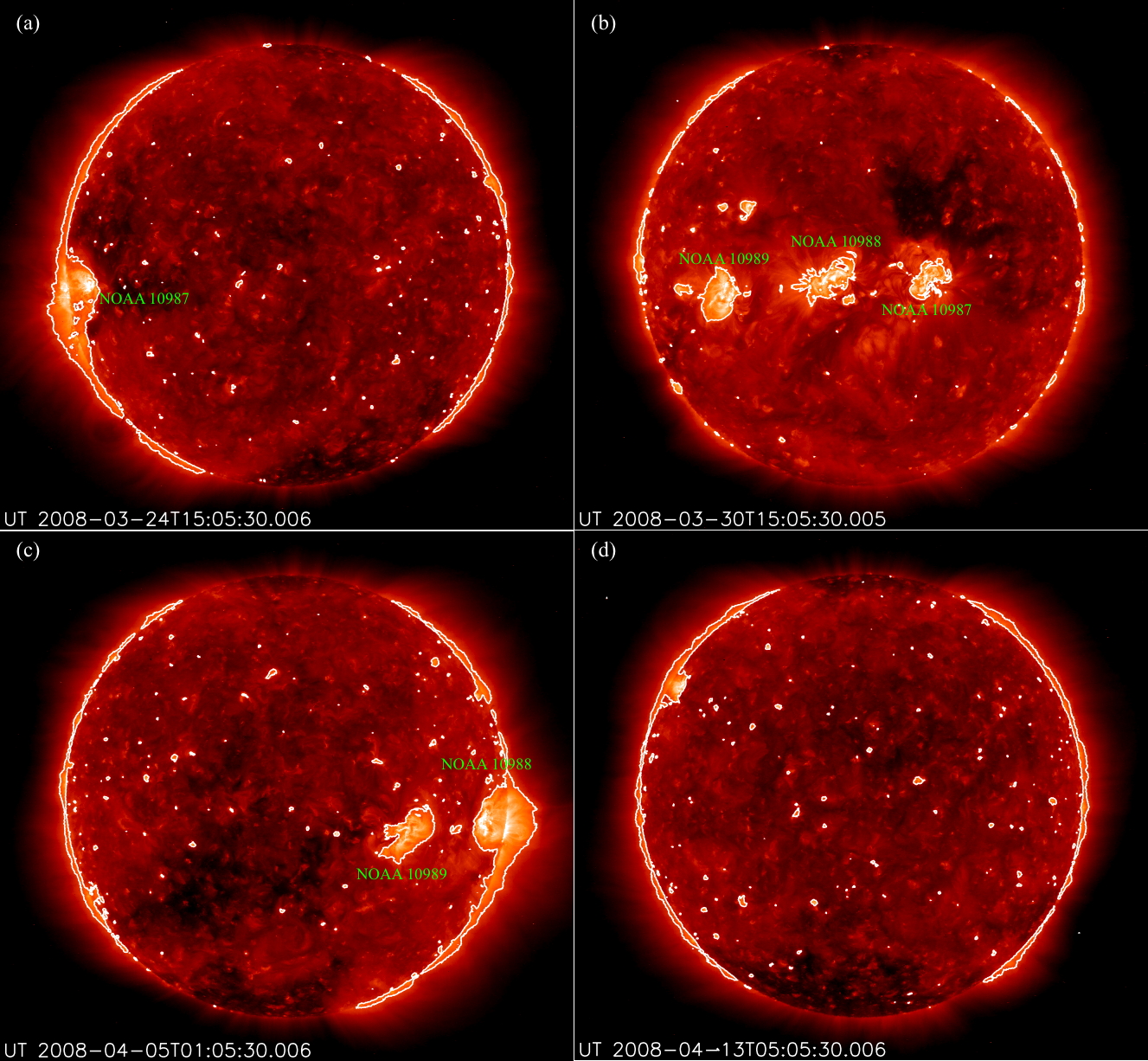}
\caption{The coronal emission images taken with EUVI/STEREO A in 195 \AA. Contours represent brightness levels at 4.5, 45.0 and 450.0 times the median disk emission. Bright patches correspond to the active regions. Active regions appear (a) on the east limb, (b) disk center, (c) the west limb and (d) the far side of the sun.  \label{images}} 
\end{center} 
\end{figure}

With EUV images, Fig. \ref{images} illustrates how the limb brightness varies as active regions are carried round by the solar rotation.  In general, coronal emission lines are optically thin. The integral along the line of sight leads to stronger emissions on east and west limbs than those on limbs near polar holes (see all panels in Fig. \ref{images}). This is simply because equatorial regions are sources of strong coronal EUV emissions due to concentrations of active regions. When sunspot groups are near the solar limb, bright emissions are enhanced from the equator  to polar regions. In panels (a) and (c), streamers outlined by contours extend from active regions to high latitudes. However, they do not cover the polar holes, indicating that these high latitude streamers are longitudinally aligned with active regions. They extend largely to high latitude through either the projection effect or by having magnetic anchors at high latitudes. When active regions are on the disk, the limb emissions are no longer intense, but, for example, the southern polar hole becomes invisible (see Panel (b)). This is because the high latitude streamers associated with the active regions appears as a large hot blob at high altitude covering the polar hole when projected onto the plane of the sky. Active regions carried to the far side of the sun produce weak limb emissions, but the southern polar hole is visible (Panel (d)). This is because the high latitude streamers connected with active regions are on the far side of the sun. They are projected to the plane of the sky, but are behind the solar disk. 
%JING - I DO NOT UNDERSTAND.  IT SHOULD NOT MAKE ANY DIFFERENCE NEAR SIDE OR FAR SIDE BECAUSE THE PLASMA IS OPTICALLY THIN.  ONLY THE  B0 EFFECT CAN MAKE ANY DIFFERENCE.
Variations in the shapes of polar holes corresponding to the projected locations of active regions on either east or west limbs were also reported by \citet{1989SoPh..119..323S} using coronal images at He I 10830 \AA. Comparing HeI 10830 images with the photospheric magnetograms, they interpreted the observations as the result of the reconnections between the polar magnetic field and the newly erupted large bipolar regions.

\subsection{Long-Lived Coronal Streamers}
\begin{figure}[t]
%\epsscale{1.2}
\begin{center}
%\plotone{lca2006339-480noaa.pdf}
\includegraphics[width=1.0\textwidth]{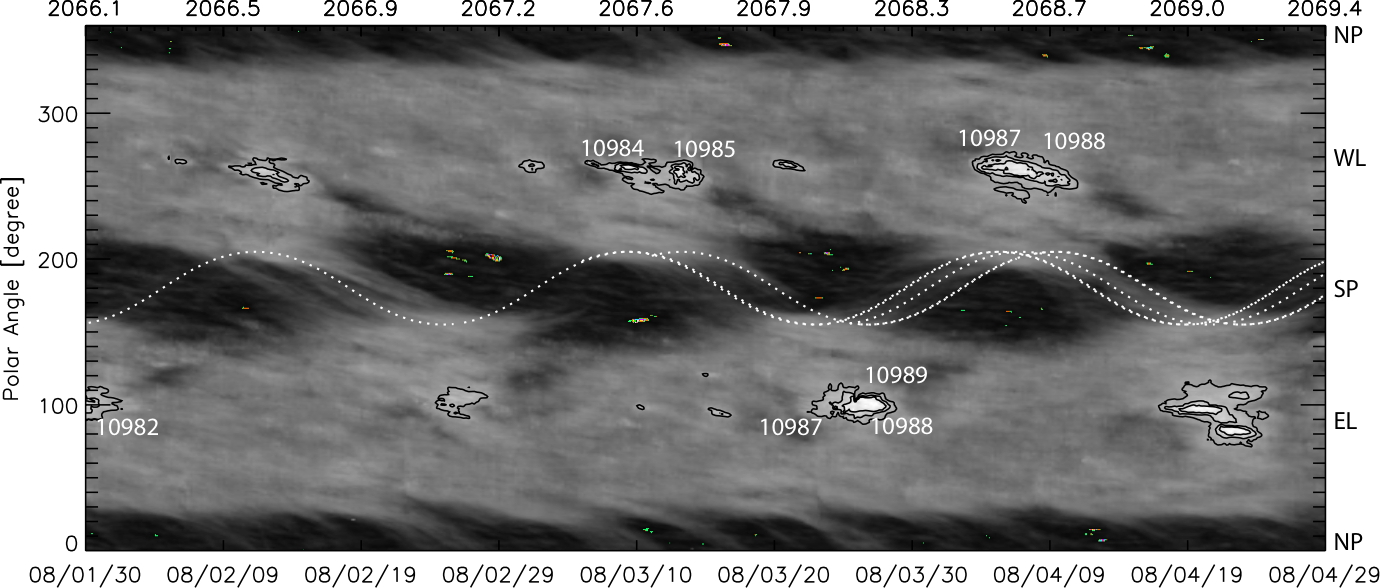}
\caption{Modeling a {\it Polar Sinusoidal Streamer} on a LSM. Sunspots and bright plage regions are outlined by contours. White dotted sinusoidal curves are calculated with Equations (\ref{latitude}) and (\ref{longitude}) given the latitudes $\delta=-65^\circ$. Numbers of curves correspond to sunspot groups marked with NOAA numbers. They shares the same rotation rates and phases as those numbered sunspots, and following a PSS. Each sinusoidal curve is plotted for 2 rotational periods.
\label{pss3}} 
\end{center} 
\end{figure}

Variations in the appearance/disappearance of the polar coronal holes and in the more equatorial limb brightness are best displayed on a so-called coronal Limb Synoptic Map (LSM) \citep{2000ApJ...539L..67L}.  The practical advantage conveyed by LSMs over conventional Carrington maps is that  variations in limb activity occurring on timescales that are very long compared to a solar rotation can be easily seen. 
%To construct a LSM, each coronal image is divided into annuli centered on the middle of the solar disk. The coronal emission is extracted around each annulus, and then plotted vertically in the LSM (c.f. Fig. \ref{pss3}). Stacked vertical lines from successive images form a LSM in which the y-axis represents polar angle around the annulus and the x-axis represents time. The 2-dimensional maps show time running from the past to the future along the x-axis, and polar angles along the y-axis, where $0^\circ$ (equivalently 360$\degr$) corresponds to the north pole, $90^\circ$ to the east limb, $180^\circ$ to the south pole, and $270^\circ$ to the west limb. 
Fig. \ref{pss3} shows a sample LSM centered on the South Pole for the period Jan 30 to Apr 29, 2008. A long-lived streamer appears as a diagonal streak crossing over the southern polar hole on three consecutive solar rotations. This pole-crossing feature is the same dense, hot plasma seen on individual images extending from active regions to  heights large enough that it is visible above the limb near the pole. 
%They cross the limb at different latitudes by projection as active regions are carried round by the solar rotation. 

If the solar spin axis were perpendicular to the ecliptic plane, polar streamers caused by projection would appear as nearly perfect sinusoidal patterns against the polar holes on LSM. In reality, the solar spin axis is inclined to the ecliptic normal by an angle  7.25${\degr}$, so that the projected tilt along the line of sight varies between $-7.25^\circ$ to $7.25 ^\circ$ through the year.  This seasonal variation in the projected tilt is responsible for an observed asymmetry in the polar streamers, such that the rising and falling sides of a polar sinusoid are not in general equally bright  \citep{2000ApJ...539L..67L}.  For example, in Fig. \ref{pss3}, the branches from the west to east limbs (the ``falling branches'',  formed when responsible active regions are on the far side of the sun) are more prominent than those from the east to west limbs (``rising branches'', formed when responsible active regions are on visible side of the sun).  This asymmetry is introduced by  the non-zero heliographic latitude of the Sun ($B_0$ fell in the range $-7.3^\circ$ to $-1.3^\circ$ viewed from STEREO A in the interval plotted in Figure \ref{pss3}). The closer is $B_0$ to zero, the more symmetric are the rising and falling branches of the long-lived streamer and the sense of the asymmetry reverses as $B_0$ changes sign. 

\begin{figure}[t]
%\epsscale{1.2}
\begin{center}
%\plotone{lca2006339-480noaa.pdf}
\includegraphics[width=0.55\textwidth]{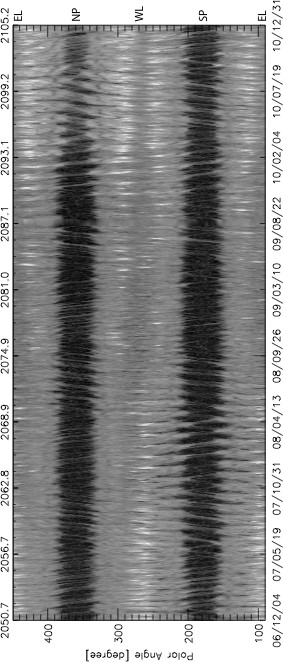}
\clearpage
\caption{
%Coronal Full Limb Synoptic Map (LSM) made with EUVI 195/STEREO A images. The full limb emissions are extracted between solar altitudes 1.045 and 1.075 \RSUN. The horizontal axis represents time from past to future (lower horizontal axis), and Carrington numbers (upper horizontal axis). The vertical direction represents the polar angle around the solar limb. $0^\circ$ is the north pole (NP); $90^\circ$ is the east limb (EL),  $180^\circ$ is the south pole (SP), $270^\circ$ is the west limb (WL). The display the entire northern polar hole, the map is repeated from the northern pole to the east limb in the polar angle. The $360^\circ$ is the north pole, and the $450^\circ$ is the east limb.
 \label{lsm}}
\end{center} 
\end{figure}

To give a sense of the life spans of long-lived coronal streamers, Fig. \ref{lsm} shows the LSM made from EUV images taken between the end of 2006 and the end of 2010.  We have plotted the total range of polar angles from $90^\circ$ to $450^\circ$  so that both polar holes (at 180$\degr$ and 360$\degr$) are displayed uninterrupted by the axes of the figure.  
The long-lived coronal streamers are seen as repetitive, inclined structures crossing the dark polar holes (marked NP for North Pole and SP for South Pole), in both hemispheres in Figure \ref{lsm}. Other prominent features of the LSMs are periodic bright patches near the equator on the east ($\sim90^\circ$) and west ($\sim 270^\circ$) limbs. They are active regions carried to the east to the west limbs by solar rotation and showing evolution in size and brightness from rotation to rotation. That they always appear as extensions of long-lived streamers indicates a physical connection between the streamers and the active regions.

%Five rectangle boxes are placed near both north and south poles on the LSM to highlight these structures. 
%They appear as periodic structures with lifetime of several solar rotations. %On an individual coronal emission image, they are large coronal streamers rooted at active regions (see Fig. \ref{image}). They are long-lived streamers associated with active regions. 
The long-lived streamers have lifetimes longer than typical sunspots because they are replenished by the continuous emergence of active regions in the photosphere. The idea was also proposed by \citet{1986SoPh..104..425S} who explained the recurrent patterns of solar magnetic field with periods 28- to 29-day. In their view, ``the patterns are caused by longitudinal fluctuations in the eruption of new magnetic flux, the transport of this flux to mid latitudes by supergranular diffusion and meridional flow''.  The long-lived coronal streamers necessarily large-scale  because they must extend high enough to be seen in projection above the polar regions in order to be detected on LSMs. It is the structured distribution of these long-lived streamers in Fig \ref{lsm} which provides the basic evidence for long-lived, non-randomly distributed coronal structures on the sun and for the underlying active longitudes of sunspots. 
%As new sunspots emerge in the vicinity of existing sunspots, coronal streamers associated with these sunspots are carried by the solar rotation, and form {\it polar sinusoidal streamers} on a LSM. 
%Because they are supported by the continuous emergence of sunspot groups, the coronal structures can persist much longer than any one sunspot, as seen by the dominance of repetitive features in Figure \ref{lsm}. 
In the present data we find numerous long-lived streamers with lifetimes $>$5 solar rotations.  The longest lasting coronal streamer was observed in the last solar minimum between 1996 April 9 and 1997 February 3, corresponding to more than 10 solar rotations \citep{2002ApJ...565.1289L}.  

\section{Sunspot Active Longitudes}
The active longitudes are the longitudinal bands from where the sunspots erupted repeatedly over a long period of time. The term is equivalent to sunspots clustering  in longitude-time space, or non-contemporaneous spot cluster.
%In Carrington longitude-time space, sunspots are clustered. 
The method to detect the spot cluster can be summarized in two basic steps: (1)  large-scale, long-lived coronal streamers on the coronal LSM are connected with sunspot groups; 
%These are normally large sunspots surviving for long period of time, or sunspots at higher latitudes giving rise to coronal streamers; 
(2) more members of a spot cluster are identified based on having Carrington longitudes in the vicinity of those spot groups already belonging to a spot cluster. 

\subsection{Sunspot Clusters by Long-lived Coronal Streamers}

The long-lived coronal streamer on the LSM can be modeled by considering the apparent sky-plane motion of a radial coronal structure as it is carried round by the solar rotation. This is equivalent to the scenario that coronal streamers are rooted at the active regions, and extend radially to high altitudes. The projected latitude, $\theta$, of such a structure as a function of time is given by: 

%\begin{eqnarray}
\begin{equation}
\theta(t)=\arctan\left(\cos B_0\frac{\tan B}{\cos L(t)}-\tan L(t)\sin B_0\right),  
\label{latitude}
\end{equation}
%\end{eqnarray}

\noindent where $B_0$ is the heliographic latitude of the solar center and $B$ is the Carrington latitude where the coronal structure is rooted.   Quantity $L(t)$ is the solar disk longitudinal position of the sunspot group as a function of time, given by

\begin{equation}
L(t)=\Omega (t-t_0) \nonumber,
\label{longitude}
\end{equation}

\noindent where $t_0$ is the time when the sunspot group crosses the east limb and $\Omega=\omega(B)-\omega_{obs}$ is the solar synodic rotation rate in which $\omega(B)$ is the sidereal rotation rate appropriate to latitude $B$. The function $\omega(B)$ is the subject of many past and continuing studies \citep{2000SoPh..191...47B,2011ApJ...730...49L}. We used the differential rotation of \citet{newton1951}, in which $\omega(B)=a+b\sin^2B$, where $a=14^\circ.38$ day$^{-1}$ and $b=-2.^\circ96$ day$^{-1}$. We adopted $\omega_{obs}=360/346$ $^\circ$day$^{-1}$ for STEREO A ($360/388$ $^\circ$day$^{-1}$ for STEREO B) to account for the drift motions of these spacecraft.  With Equations (\ref{latitude}) and (\ref{longitude}), a radial plasma structure above an active region describes a sinusoidal curve on a LSM.  This is shown in the figure \ref{pss3} in which a long-lived streamer is traced by a number of {\it sinusoidal curves} (dotted curves). The long-lived streamer varies in phase with the NOAA-numbered sunspots, which jump from the east to the west limbs as they are carried round by solar rotation. Dotted sinusoidal curves have the same phases and periods as sunspot groups marked with NOAA numbers, but are rooted at latitude $B=-65^\circ$. This indicates that the active regions and the high latitude streamers are longitudinally locked  and rigidly connected. These sunspot groups are members of a sunspot cluster because their corresponding polar sinusoidal curves collectively fit to a long-lived coronal streamer.

Sunspots corresponding to a long-lived coronal streamer in Fig. \ref{pss3} have the following properties: (1) they appear on either the east or the west limb; (2) they give rise to large-scale streamers projecting to high latitudes; (3) their corresponding {\it polar sinusoidal curves} calculated with Equations (\ref{latitude}) and (\ref{longitude}) on a LSM unambiguously match a long-lived streamer. About 50\% of sunspots in a cluster are identified in this step. 
%Some sunspots which emerge in the vicinity of members of clusters, but do not satisfy these conditions, are likely left out as members of a non-contemporaneous spot cluster. 

\subsection{Spot Cluster Membership based on Longitudinal Preference}
If the sun rotated rigidly at the Carrington rotation rate, $\omega_{cr}$, we would see sunspot groups emerging in the vicinity of a fixed Carrington longitude one rotation after another. The long-lived streamers on a LSM would have the same period as the Carrington rotation period. In reality, the Carrington longitude of an active region drifts in time due to differential rotation:

\begin{equation}
\lambda(t)=\lambda_0(t_0)+[\omega-\omega_{cr}] (t-t_0)
\label{clng}
\end{equation}

\noindent where $\lambda_0$ is the emerging sunspot Carrington longitude at $t_0$;  $\omega$ is the rotation rate of the sunspot; $\omega_{cr}=14.18^\circ$ day$^{-1}$ is the sidereal Carrington rotation rate. 

\begin{figure}[t]
\epsscale{1.0}
\begin{center}
%\plotone{fig3.jpg}
\includegraphics[width=1.0\textwidth]{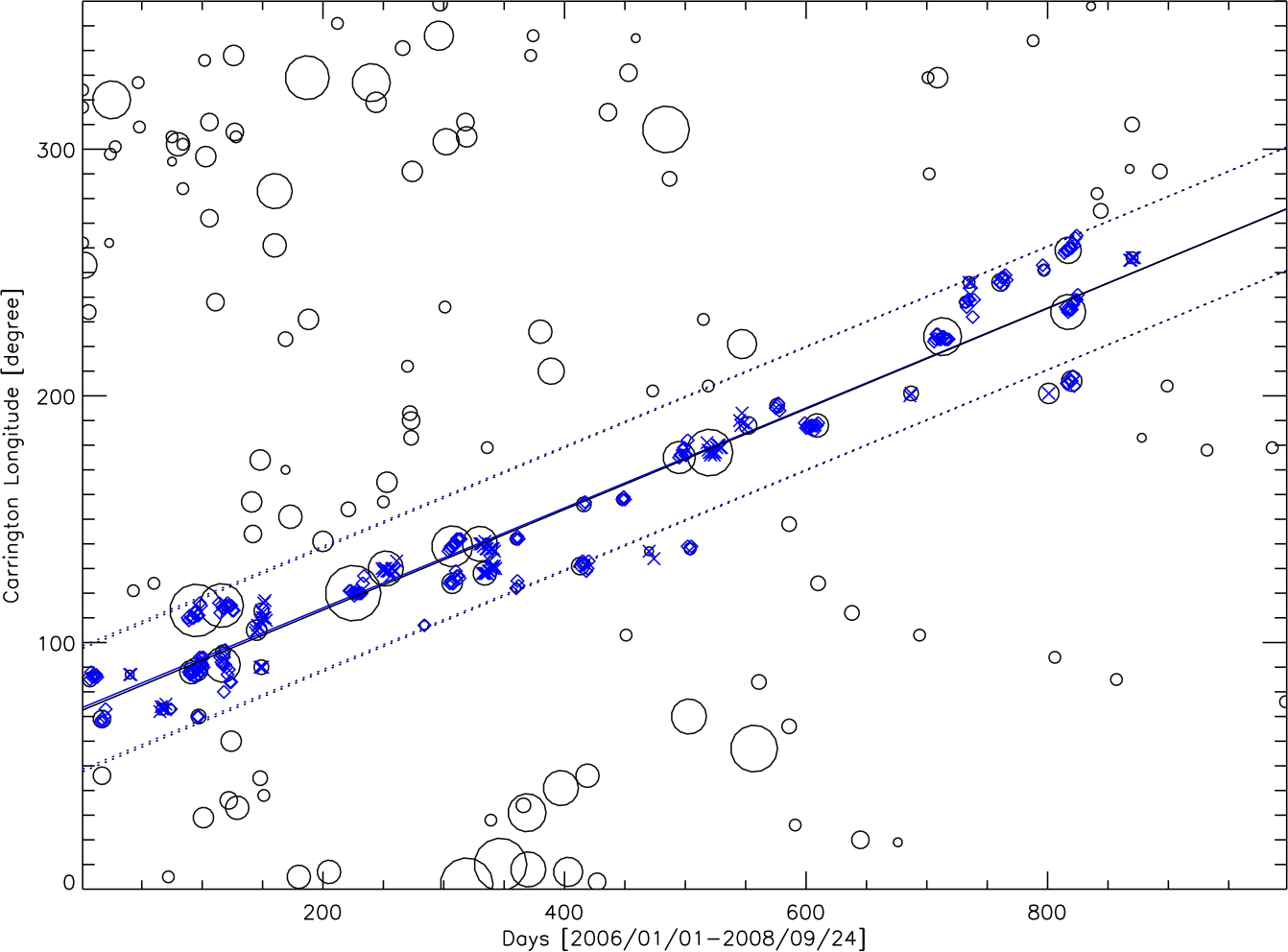}
\caption{Carrington longitudes of sunspot groups as functions of time. All sunspot groups emerging between 2006 January 1 (DOY 1) and 2008 September 24 (DOY 998) are plotted with circles whose radii  are proportional to the sunspot radii , $\sqrt s/\pi$, where ``$s$'' is the sunspot group area recorded in SRS. The sunspots in blue colors are members of a non-contemporaneous cluster. The sunspots with blue ``$\diamond$'' are identified by a long-lived {\it polar sinusoidal streamer}. A linear regression fit of these sunspots is plotted in a blue straight line. The longitudinal range of $\pm 25^\circ$ about the line is defined by blue dotted lines.  The blue ``$\times$'' represent added sunspots to the existing spot cluster because they distribute within the two dotted blue lines. A black solid straight line is the linear regression fit to all members of the spot cluster. Note that there is only a little difference between the black and blue lines.%JING - SHOULD THE SPOTS ALSO HAVE PURPLE CIRCLES?
 \label{sc3}} 
\end{center} 
\end{figure}

Fig.\ref{sc3} is an example of how sunspot groups are added  to an existing non-contemporaneous sunspot cluster.  All sunspot groups emerging between 2006 January 1 and 2008 September 24 (DOY 998) are plotted with circles on the Carrington longitude versus time plot. The radii of the circles are proportional to the sunspot group radii, $\sqrt s/\pi$, where ``$s$'' is the sunspot group area obtained from the SRS. The sunspot groups colored with blue ``$\diamond$'' are those identified with a long-lived streamer in step 1. Essentially all large spots in the cluster are identified in this way.  Small spots evidently do not produce plasma structures tall enough to be seen in projection over the poles, and therefore do not form polar streamers. A linear regression fit of blue $\diamond$ spots with $\pm 25 ^\circ$ border lines define the range of Carrington longitudes of a spot cluster. Sunspot groups falling inside this longitudinal range, but are not previously recognized as the members of the cluster in step 1, are marked with blue ``x''. These are new members to the existing spot cluster. About 50\% of the spots in a  cluster are identified in this way. The black solid line is a linear regression fit to all members of the spot cluster.  The properties of the spot clusters do not depend on whether we use only those sunspots identified using the polar streamers as proxies or the totality of sunspots identified in steps (1) and (2), above.  

\section{Results and Discussions}
\begin{figure}[t]
\begin{center}
%\plotone{sc_fits.pdf}
\includegraphics[width=1.0\textwidth]{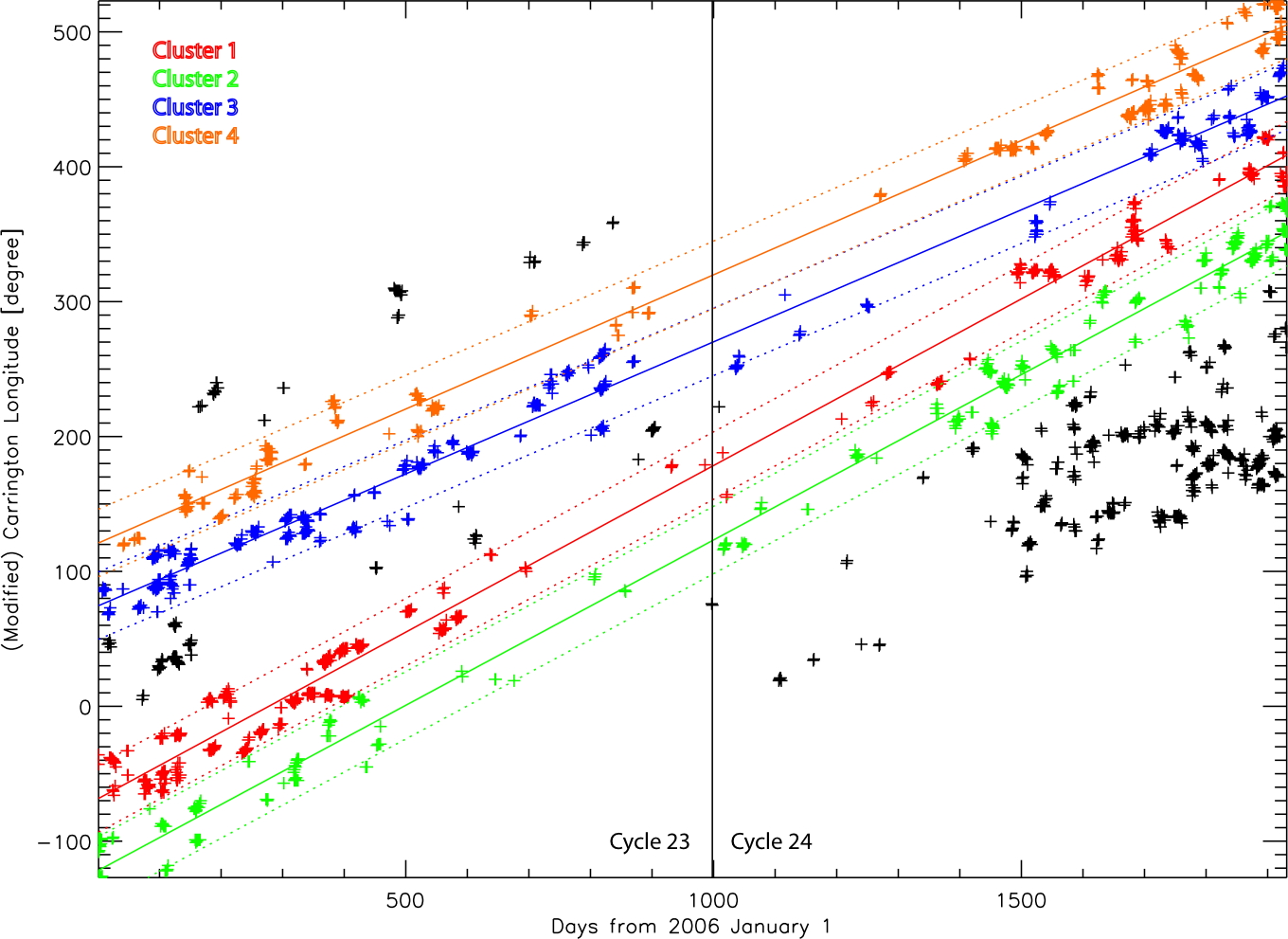}
\caption{Carrington longitudes of sunspots as function of time between 2006 January 1 and 2011 April 11. Members of non-contemporaneous clusters are plotted in colored ``+''. The linear regression fits of Carrington longitudes and $\pm 25^\circ$ of the fits  are drawn with colored straight and dotted lines. Color schemes are used to represent sunspot groups associated with respective non-contemporaneous sunspot clusters:  red, green, blue and orange represent spot clusters 1, 2, 3 and 4, respectively.  The Carrington longitudes of some sunspots have been added or subtracted by $360^\circ$ to avoid artificial roll over from $360^\circ$ to $0^\circ$ within a cluster. The vertical straight line is drawn at DOY=998 (2008 September 24) separates the cycle 23 from 24. The sunspot groups found having no clear affiliation with any clusters at the present time are marked with black ``+''.  \label{4sc}} 
\end{center} 
\end{figure}

Figure \ref{4sc} plots all sunspot groups recorded between 2006 January 1 and 2011 April 15. More than 75\% of numbered sunspots during this period are members of four spot clusters, which are plotted in the figure with differently colored symbols. Additional spot groups having no affiliation with any cluster are marked with black ``+''.  (We note that, although those spots cannot be associated with a cluster based on any repetitive coronal structures, their collective longitudinal distribution does appear non-random in Fig.\ref{4sc}). In the current discussion, we will focus on the four sunspot clusters. In the Figure, we have added or subtracted $360^\circ$ to/from the Carrington longitudes of some sunspots of clusters in order to avoid artificial roll-over  as a spot longitude drifts upwards from below $360^\circ$. DOY 998 (2008 September 24) is designated the end of cycle 23 based on the time when major sunspots first appeared at high latitudes (see the vertical line).  All four spot clusters contain sunspot members emerging in both cycles 23 and 24.  In other words, the clusters persist across the boundary between solar activity cycles.

From Fig. \ref{4sc}, a number of properties of a spot cluster are drawn and they are summarized in Table \ref{sc}.  In the Table, the median Carrington latitude, $\bar B$,  is the median of the latitudes of all the sunspots within a cluster.  From the linear regression fits of sunspot Carrington longitudes against time within a spot cluster (see straight lines going through each cluster in Fig.\ref{4sc}), we obtain $\lambda_0$, the Carrington longitude at time $t_0=0$. The slope of the linear regression fits (c.f. Equation \ref{clng}) gives, $k=\omega-\omega_{cr}$ ($\omega_{cr}$ is the Carrington rotation rate), and each spot cluster corresponds to a constant $\omega$.  The surface sidereal rotation rates, $\omega(\bar B)$, are calculated from the differential rotation of \citet{1970SoPh...12...23H}, which measures the surface rotation rates using the spectral line shift data recorded at Mt. Wilson. Finally,  synodic rotation periods for Earth are given in the last column, derived from the cluster common rotation rates, $\omega$.
%JING - BUT EARLIER DIDNT YOU SAY THAT YOU USED A DIFFERENT PAPER FOR OMEGA_C? NEWTON AND NUNN, IN FACT.  I"M CONFUSED.

\begin{figure}[t]
\epsscale{1.0}
\begin{center}
%\plotone{fig3.jpg}
\includegraphics[width=0.5\textwidth]{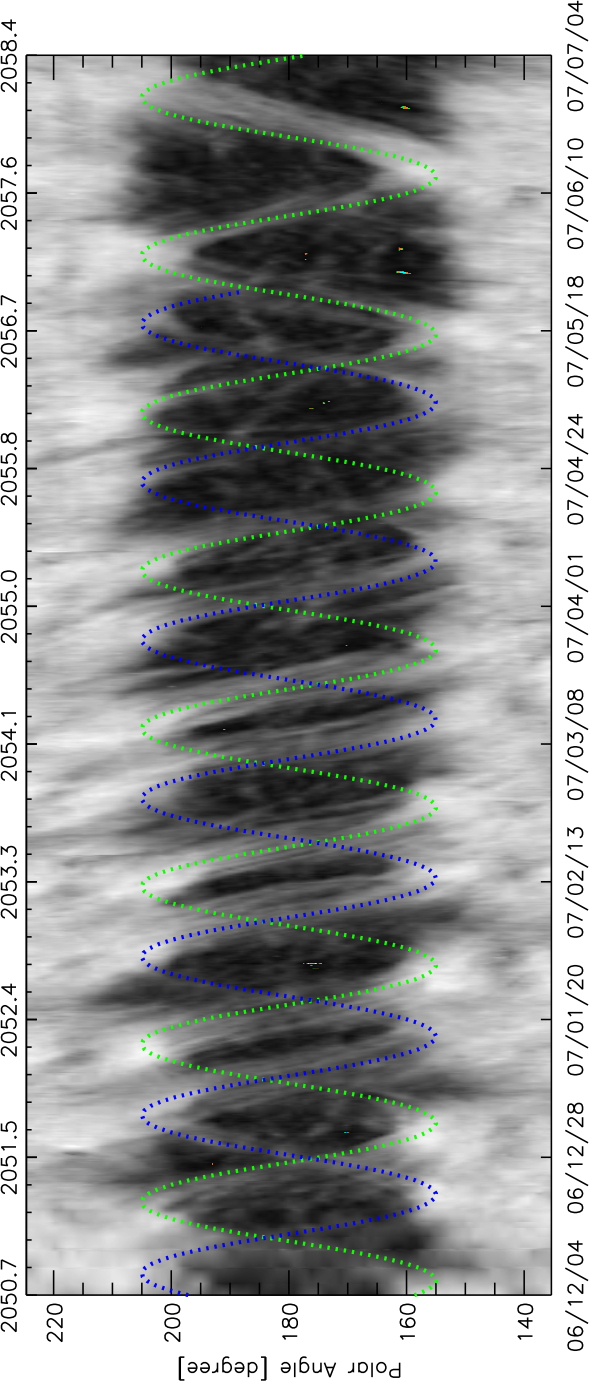}
\caption{Two long-lived coronal streamers are traced by sinusoidal curves against southern polar hole ($180^\circ$) between 2006 December 4 and 2007 August 23 (the vertical axis varies from $135^\circ$ to $225^\circ$). They correspond to clusters 2 and 3. The green and blue dotted curves are calculated with the equations (\ref{latitude}) and(\ref{longitude}). They are rooted at $B=-65^\circ$ and have periods 26.88 and 26.99 [days] ($=360^\circ/(\omega-\omega_{obs})$. $\omega$ is the cluster rotation rate, and $\omega_{obs}=360^\circ/346$ for STEREO A), respectively.   \label{crr}} 
\end{center} 
\end{figure}

What Fig. \ref{4sc} has revealed is that a spot cluster corresponds to a single rotation rate, $\omega$. The rotation rate also describes the period of a long-lived coronal streamer on a LSM. This is shown in Fig. \ref{crr} in which the dotted curves calculated with equations (\ref{latitude}) and (\ref{longitude}) trace two long-lived coronal streamers. They correspond to spot clusters 2 and 3 having $\omega=14.43^\circ$ day$^{-1}$ and $14.38^\circ$ day$^{-1}$, respectively, but are rooted at $-65^\circ$ latitude. As we have discussed before, these long-lived streamers are connected with more equatorial active regions. Although two long-lived streamers have slight different rotation periods, 26.88 and 26.99 days, respectively, this may not be detected when they are used as tracers to measure the coronal rotation. The fact that each sinusoidal curve follows a single period indicates that corona likely appears to rotate constantly and rigidly when large-scale streamers are used as tracers. This is consistent with past observations using the long-lived streamers \citep{1974SoPh...34....3A,1984ApJ...287..959F,1989ApJ...336..454S}.   They concluded  that the corona rotates more rigidly and faster than the photosphere. On the other hand, the short-lived coronal activity tends to show the same differential rotation as in the photosphere \citep{1974SoPh...34....3A,1995SoPh..160....1I,2009ApJ...697..980H}. The close relationship between the coronal rotation and photospheric magnetic flux eruptions is proposed by \citet{1988ApJ...327..427W} using spherical harmonic analysis and numerical simulations. Our observation confirms that the solar corona largely reflects the photospheric emerging flux.  
%Active longitudes in the photosphere directly result in the rigid rotation of the coronal streamers.

%JING - I CANNOT UNDERSTAND THE PREVIOUS PARAGRAPH AT ALL.

That spot cluster rotation rate is shared by all members indicates the existence of a layer with a constant speed on the sun, from where sunspots of a cluster are rooted.  From the GONG 
%JING - I THOUGHT THE REFEREE POINTED OUT THAT GONE IS AN ERROR.
and MDI data, helioseismology reveals that the solar rotation rate is a function of the depth as well as the latitude. The rotation rate at the equator reaches a maximum (470 nHz) just below the surface ($\sim 0.95$ \RSUN), then decreases with increasing depth. The rotation becomes independent of latitude at 0.70 \RSUN~(the base of the convection zone), where the rate is 440 nHz ($\sim 13.69^\circ$ day $^{-1}$) (RLS inversion) \citep{1996Sci...272.1300T}. The cluster rotation rates ($\sim 460$ nHz in Table \ref{sc}) are faster than the bottom of the convection zone by 5\%.  On average, the cluster rotation rates are also $\sim$5\% faster than the surface rates determined by Doppler measurements \citep{1970SoPh...12...23H} (see the column $\omega(\bar B)$ in Table \ref{sc}).   This suggests that a majority of sunspot groups are rooted beneath the photosphere and half-way to the bottom of the convection zone. The rotation rate of the cluster pair 1 \& 2 ($14^\circ.43\pm 0.01$ day$^{-1}$)  differs from that of the pair 3 \& 4 ($14^\circ.38\pm 0.01$ day$^{-1}$) by $0.05^\circ$ day$^{-1}$. Although small, this difference  can be significant, amounting to about 200$\degr$ on the solar cycle timescale. A complete image of active longitudes throughout the whole solar cycle is needed to understand how the spot clusters evolves longitudinally. 
%JING - THE LAST SENTENCE DANGLES.....SO WHAT?

\begin{figure}[t]
%\epsscale{0.95}
\begin{center}
\includegraphics[width=1.0\textwidth]{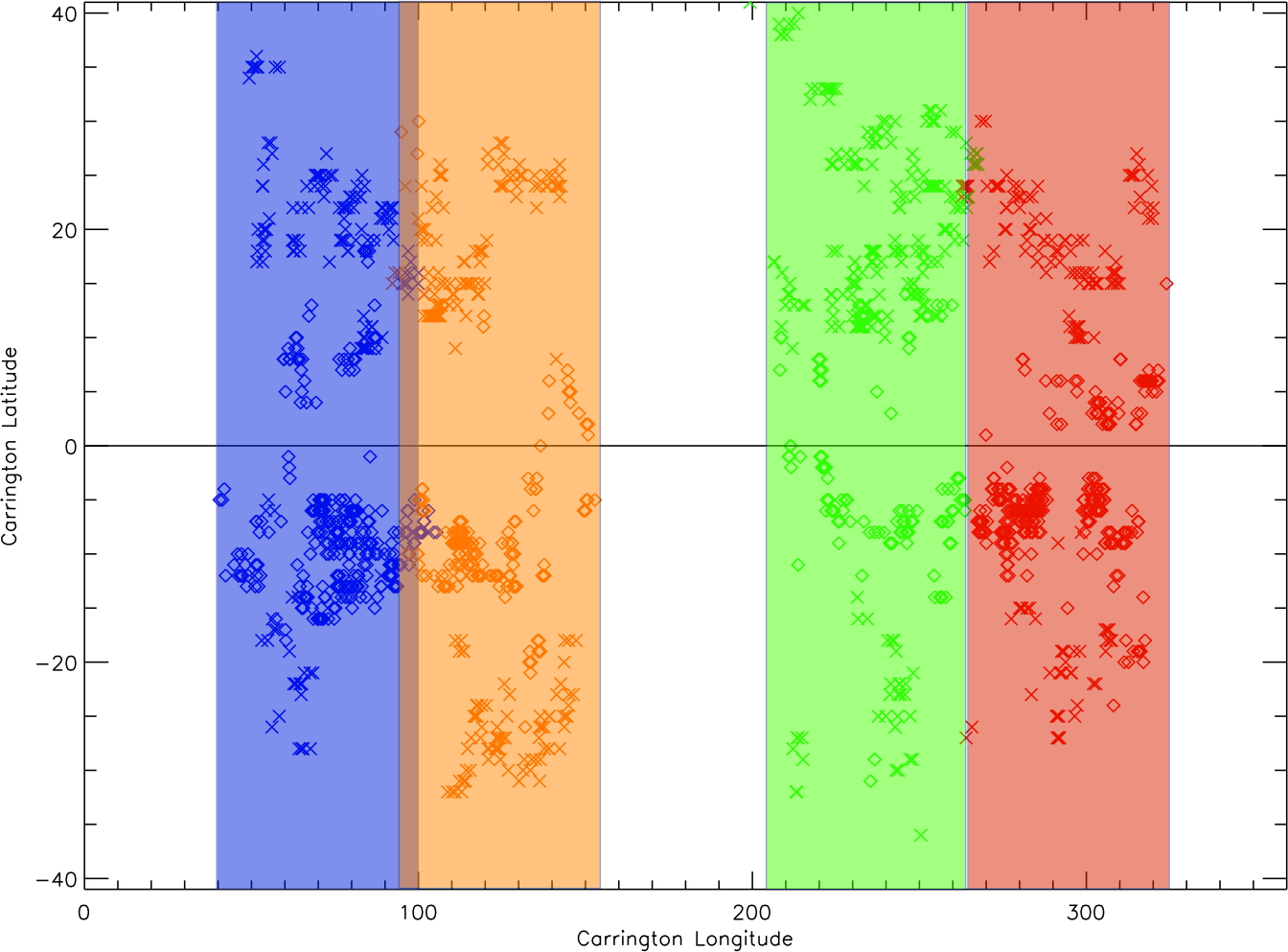}
\caption{Sunspot longitudinal distribution. All sunspot groups affiliated with non-contemporaneous spot clusters between 2006 January 1 and 2011 April 11 are plotted with their Carrington latitudes against longitudes. Their longitudes have been shifted by $(\omega-\omega_{cr})(t-t_0)$, where $t_0$ is 2006 January 1, as if they all emerged on $t_0$. The color scheme is used to distinguish spot clusters and has the same meaning as used in Fig. \ref{4sc}. Within each spot cluster, sunspot groups emerging before DOY 998 (2008 September 24) are plotted with ``$\diamond$'', those emerging after DOY 998 are plotted with ``$\times$''. Each cluster is further distinguished with colored transparent rectangles indicating the longitudinal range of the cluster. The solid horizontal line shows the equator.\label{clng_correct}} 
\end{center} 
\end{figure}

Fig. \ref{4sc} also reveals that active longitudes on the sun persist regardless of solar cycle or latitude.  
%JING - IS THIS RIGHT?  I CHANGED IT
Fig. \ref{clng_correct} shows the sunspot groups of the four clusters as if they all emerged on DOY = 1 (2006 January 1). The Carrington longitudes of sunspots are shifted by the quantity $[\omega-\omega_{cr}](t_0-t)$ within each cluster; $t_0=$2006 January 1,  and $t$ is the time of a sunspot emergence. The color scheme in the figure has the same meaning as those in Fig.\ref{4sc}. To distinguish the (low latitude) sunspot groups emerging in the end of cycle 23 from the (high latitude) sunspot groups emerging in new cycle 24,  the cycle 23 spots are represented by ``$\diamond$'' symbols while those of cycle 24 are represented by ``$\times$'' symbols. The four clusters of sunspots clearly occupied four distinct longitudinal bands, each of them spanning $\sim$60${\degr}$ in longitude. Although spot groups emerging in the new cycle occupy higher latitudes than those in the old cycle, consistent with the general trend of the Butterfly diagram, there is no evident change in the distribution of longitudes between spots appearing in the old and the new activity cycles.  Figure (\ref{clng_correct}) shows that the longitudinal clustering survives, indicating that whatever underlying magnetic structure causes the longitudinal clumping is not destroyed by the emergence of a new activity cycle.
%Among four spot clusters, the cluster pair 1 and 2,  the pair 3 and 4 are closely longitudinally distributed. The initial Carrington longitudes represented by $\lambda (t_0)$ are only $\sim45 ^\circ$ apart between cluster pairs, while the average longitudinal separation between the two pairs is $\sim 180^\circ$. 

\begin{figure}[t]
%\epsscale{0.95}
\begin{center}
\includegraphics[width=1.0\textwidth]{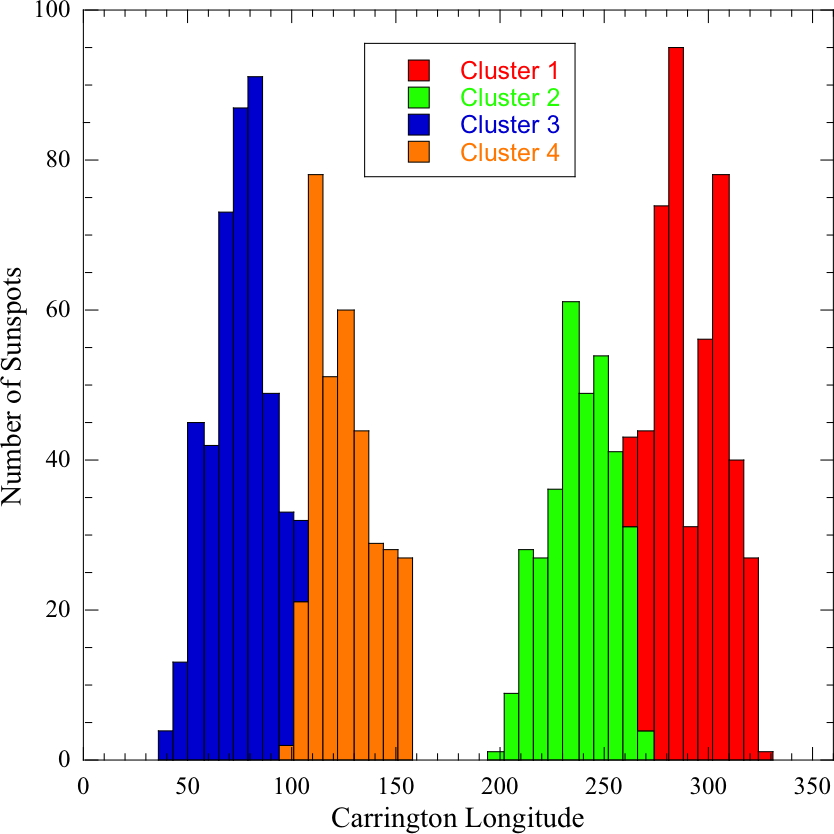}
\caption{Sunspot longitudinal histogram. The sunspot longitudes have been shifted as in Figure \ref{clng_correct}, as if all sunspots emerged on 2006 January 1.\label{histogram}} 
\end{center} 
\end{figure}

Fig. \ref{histogram} further demonstrates the distribution of the sunspot longitudes within the four clusters, again plotted as if the spots all emerged on DOY =1 (2006 January 1). The distributions are clearly non-uniform, with preferred bands centered at roughly $90^\circ$ for the cluster pair 3 \& 4 and $270^\circ$ for the cluster pair 1 \& 2, respectively. Both bands have longitudinal range $\sim 100^\circ$.  The centers of the two massively occupied longitudinal bands are separated by $\sim180^\circ$.

\section{Summary}
We use EUV images from a $>5$ year (2006 January 1 and 2011 April 15) period around the recent solar activity minimum to identify long-lived coronal streamers.  Comparing coronal streamers on Limb Synoptic Maps with photospheric sunspot records, we find that:

\begin{enumerate}

\item The EUV coronal streamers persist for many  solar rotations and are non-randomly distributed in time and/or solar longitude. These long-lived structures are best detected on full coronal Limb Synoptic Maps. 

\item The long-lived, large-scale coronal streamers on Limb Synoptic Maps reflect non-random longitudinal distributions of photospheric sunspot groups. They are long-lived because they are sustained by sunspot groups emerging in the vicinity of a longitude, but at different times on the sun.  
%They reveal the existence of sunspot active longitudes. 

\item Sunspots emerge clustered in longitude-time space. Lifetimes of these spot clusters are comparable to the 5-year time coverage of the current study.  More than 75\% of numbered sunspots in the period of study are members of these four spot clusters. 

\item  Sunspot clusters straddle the boundary between activity cycles 23 and 24. 

\item Members of spot clusters share common rotation rates which are faster by $\sim$5\% than both the surface and the bottom of the convection zone. The implication is that members of spot clusters are rooted in the convection zone.

\item The four spot clusters contain sunspots from both hemispheres at all latitudes. Heavily occupied longitudinal regions are characterized with two meridional bands each $\sim$100$\degr$ wide and separated by $\sim$180$\degr$. 
%This is consistent with independent reports of ``active longitudes'', suggesting an unmodelled asymmetry in the solar dynamo.

%\item When large-scale, long-lived coronal streamers are used as tracers, we anticipate that corona appears to rotate constantly and rigidly.
\end{enumerate}

\acknowledgments
The author would like to thank Dr.~Jean-Pierre W\"{u}elser at LMSAL for helping with the large amount of EUVI data retrieval, and Dr. Roger Ulrich for discussions about the solar dynamo, and recommendations on literature. She appreciates comments and encouragements from colleagues who read the manuscript. She is indebted for the critical comments made by the second referee, as well as his/her patience and suggested references. They help to clarify the descriptions of the paper. She is grateful for the tireless support of David Jewitt who made numerous critical comments, which greatly improved the manuscript. The STEREO/SECCHI/EUVI data used here are produced by an international consortium of the Naval Research Laboratory (USA), Lockheed Martin Solar and Astrophysics Lab (USA), NASA Goddard Space Flight Center (USA), Rutherford Appleton Laboratory (UK), University of Birmingham (UK), Max-Planck-Institut f\"{u}r Sonnensystemforschung (Germany), Centre Spastiale de Li\`{e}ge (Belgium), Institut d'Optique Th\'{e}orique et Applique (France) and Institut d'Astrophysique Spatiale (France). EIT/SOHO is a joint ESA-NASA program.

%$\omega=a+b\sin^2\bar B+c\sin^4 \bar B$ where $a=13^\circ.76$ day$^{-1}$; $b=-1^\circ.74$ day$^{-1}$; and $c=-2^\circ.19$ day$^{-1}$

\begin{deluxetable}{ccccccc}
\tablenum{1}
\tabletypesize{\scriptsize}
\tablecolumns{7}
\tablecaption{Properties of  Non-contemporaneous Sunspot Clusters\label{sc}}
\tablewidth{0pt}
\tablehead{
\colhead{Cluster\tablenotemark{(1)}} &
\colhead{Time Duration} &
\colhead{Carrington} &
\colhead{Carrington} &
\multicolumn{2}{c}{Sidereal Rate [$^\circ$day$^{-1}$]} &
%\multicolumn{2}{c}{Synodic Period [days]} \\
\colhead{Synodic Period [days]} \\
%\colhead{Synodic Period\tablenotemark{(7)}} \\

\colhead{} &
\colhead{} &
\colhead{$\bar B$\tablenotemark{(2)}} &
\colhead{$\lambda_0(t_0=0)$\tablenotemark{(3)}} &
\colhead{$\omega$\tablenotemark{(4)}~~~~(nHz)} & 
\colhead{$\omega(\bar B)$\tablenotemark{(5)}} &
\colhead{P($\omega$)\tablenotemark{(6)}} 
%\colhead{P($\omega(\bar B)$\tablenotemark{(7)}}
%\colhead{[days]}
}
\startdata
1 (471)& 2006/01/01 to 2011/04/15 &$~8^\circ$ & $291.4^\circ$ & 14.43 (464) & 13.73 & 26.77 \\
2 (356)& 2006/01/01 to 2011/04/15 & $15^\circ$ & $238.1^\circ$ & 14.43 (464)& 13.64 & 26.78 \\
3 (440) & 2006/01/07 to 2011/04/10 & $11^\circ$ & $~74.4^\circ$ & 14.38 (462)& 13.69&26.88\\
4 (334)& 2006/02/10 to 2011/04/08 & $13^\circ$ & $120.9^\circ$ & 14.38 (462)& 13.67 & 26.87 \\
%5 & 2009/07/04$\sim$2010/07/27 & $22^\circ$ & $290.4^\circ$ & 13.8718& 28.0653\\
%6 & 2009/11/16$\sim$2010/10/12 & $23^\circ$ & $285.2^\circ$ & 13.9138 & 27.9648 \\
\enddata
\tablenotetext{(1)}{Numbers in parentheses are the total sunspot numbers in each cluster.}
%\tablenotetext{(1)}{Median Carrington latitude, $\bar B$, and longitude, $\bar \lambda$, of the sunspot groups within each spot cluster. $\bar\lambda$ is the median Carrington longitude after the Carrington longitudes are shifted by $(\omega-\omega_{cr})\times (t_0-t)$ with $t_0$ 2006 January 1 within a cluster.}
\tablenotetext{(2)}{Median Carrington latitude of the spots within each cluster.}
\tablenotetext{(3)}{Carrington longitude at $t_0=0$ in the linear regression fit for each spot cluster.}
\tablenotetext{(4)}{Common sidereal rotation rates of spot clusters. They are derived from the slope of the linear regression fits, $k=\omega-\omega_{cr}$, where $\omega_{cr}=14.1844^\circ$ day$^{-1}$. The uncertainty is $\pm 0.01$ and is measured as $k/\sqrt{n}$, where $k$ is the linear regression fit slope, and $n$ is the number of sunspots in the fit. The rotation rates in nHz are listed in the brackets.}
\tablenotetext{(5)}{Surface sidereal rotation rate at the latitude ($\bar B$) from the formula of \citet{1970SoPh...12...23H}. $\omega=a+b\sin^2\bar B+c\sin^4 \bar B$ where $a=13^\circ.76$ day$^{-1}$; $b=-1^\circ.74$ day$^{-1}$; and $c=-2^\circ.19$ day$^{-1}$.}
\tablenotetext{(6)}{The synodic period of the spot cluster viewed from Earth. It is resulted from the spot cluster common rotation rate by the relation $P(\omega)=360^\circ/(\omega-\omega_{earth})$, and $\omega_{earth}=360^\circ/365.25$ day$^{-1}$. }
%\tablenotetext{(7)}{The synodic period related with the surface rotation rate, $P(\omega(\bar B))=360^\circ/(\omega(\bar B)-\omega_{earth})$. }
\end{deluxetable}

\clearpage

\clearpage
\clearpage

\textbf{Fig. 3} Coronal Full Limb Synoptic Map (LSM) made with EUVI 195/STEREO A images taken between December 2006 and December 2010. The full limb emissions are extracted between solar altitudes 1.045 and 1.055 \RSUN. The horizontal axis represents time from past to future (lower horizontal axis), and Carrington rotation numbers (upper horizontal axis). The vertical direction represents the polar angle around the solar limb with $0^\circ$ (and $360^\circ$) being the north pole (NP), $90^\circ$ the east limb (EL),  $180^\circ$ the south pole (SP) and $270^\circ$ the west limb (WL).  The polar angle is plotted beyond the 0$\degr$ to 360$\degr$ range to present both polar regions without interruption.
\clearpage

\clearpage

\clearpage

\clearpage

%JING
%JING - THE DOTS IN FIGURE 5 ARE TOO SMALL TO SEE AND SHOULD BE BIGGER, OR A DIFFERENT SHAPE.  

\clearpage

\end{document}